\documentclass[fleqn,10pt]{wlscirep}
\usepackage[utf8]{inputenc}
\usepackage[T1]{fontenc}
\title{Cold-atom sources for the Matter-wave laser Interferometric Gravitation Antenna (MIGA)}

\author[1,*]{Quentin Beaufils}
\author[1]{Leonid A. Sidorenkov}
\author[1]{Pierre Lebegue}
\author[1]{Bertrand Venon}
\author[1]{David Holleville}
\author[1]{Laurent Volodimer}
\author[1]{Michel Lours}
\author[2]{Joseph Junca}
\author[2]{Xinhao Zou}
\author[2]{Andrea Bertoldi}
\author[3]{Marco Prevedelli}
\author[2]{Dylan O. Sabulsky}
\author[2]{Philippe Bouyer}
\author[1]{Arnaud Landragin}
\author[2]{Benjamin Canuel}
\author[1]{Remi Geiger}

\affil[1]{LNE--SYRTE, Observatoire de Paris, Universit{\'e} PSL, CNRS:UMR 8630, Sorbonne Universit{\'e}, 61 avenue de l'Observatoire, F--75014 Paris, France}
\affil[2]{LP2N, Laboratoire Photonique, Num{\'e}rique et Nanosciences, Universit{\'e} Bordeaux--IOGS--CNRS:UMR 5298, rue F. Mitterrand, F--33400 Talence, France}
\affil[3]{Dipartimento di Fisica e Astronomia, Universit{\`a} di Bologna, Via Berti-Pichat 6/2, I-40126 Bologna, Italy}

\affil[*]{quentin.beaufils@obspm.fr}

\begin{abstract}
The Matter-wave laser Interferometric Gravitation Antenna (MIGA) is an underground instrument using cold-atom interferometry to perform precision measurements of gravity gradients and strains. Following its installation at the low noise underground laboratory LSBB in the South-East of France, it will serve as a prototype for gravitational wave detectors with a horizontal baseline of 150 meters. Three spatially separated cold-atom interferometers will be driven by two common counter-propagating lasers to perform a measurement of the gravity gradient along this baseline. This article presents the cold-atom sources of MIGA, focusing on the design choices, the realization of the systems, the performances and the integration within the MIGA instrument. 
\end{abstract}
\begin{document}

\flushbottom
\maketitle
%
%
\thispagestyle{empty}

\section{Introduction}
\label{intro}
Gravitational wave astronomy requires the development of detectors addressing the various frequency bands of interest to the astrophysics and fundamental physics communities~\cite{Sesana2016,auger2017}. Besides the current ground-based laser interferometers LIGO~\cite{Aasi2015}, VIRGO~\cite{Acernese2014} and  KAGRA~\cite{Akutsu2019} operating in the $\sim 10$~Hz to 2~kHz frequency band, and the Pulsar timing array probing the nHz region of the spectrum, the space-based laser interferometer LISA is planned to be launched in 2034 to probe the $\sim 1-100$~mHz frequency band. Beyond these detectors, several instruments are being designed to probe the intermediate frequency band $0.1-10$~Hz on the ground \cite{ET-design,shimoda_torsion-bar_2020} or in space \cite{DECIGO2017,Lacour2018}.
Ground based cold-atom interferometers represent a promising candidate solution owing to their large immunity to seismic noise \cite{Dimopoulos2008} and their potential to reject the gravity gradient noise to a sufficient extent \cite{Chaibi2016} in that frequency band. 

\begin{figure}[h!]
\centering
\includegraphics[width=0.9\linewidth]{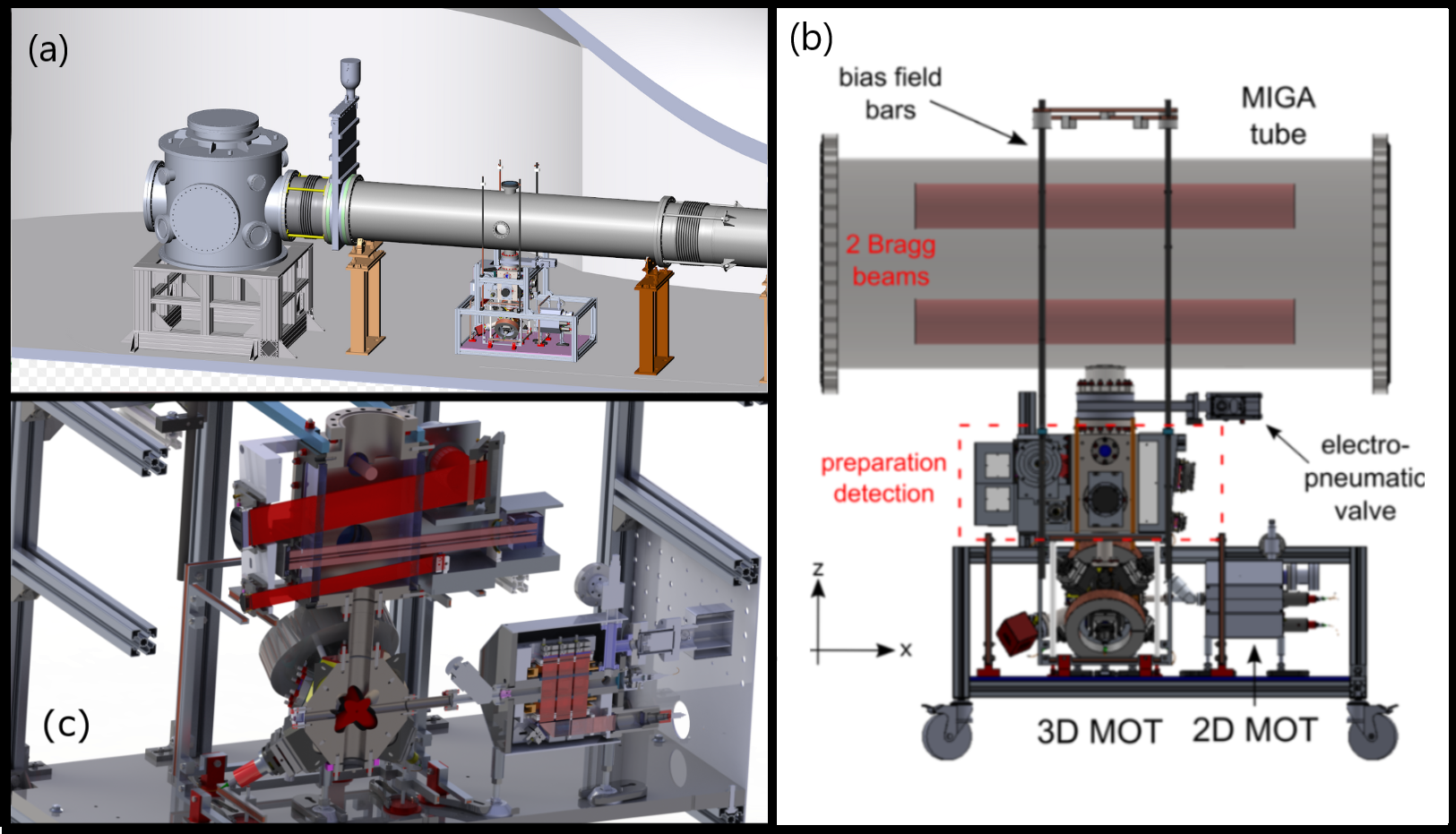}
\caption{ Overview of the MIGA instrument. (a) CAD view of the turret at the extremity of one of the two orthogonal 150-m long vacuum tubes of MIGA; one cold-atom system is shown on the right without its magnetic shields. The vertical panels represent valves to isolate the tubes from the chamber hosting the optics. (b) CAD drawing of the cold-atom source with a section of the vacuum tube (diameter: 500~mm); the two red cylinders represent the Bragg beams used to drive the atom interferometer and separated by $30.6$~cm. The distance from the ground to the bottom of the vacuum tube is about 1 meter.  (c) Sectional view of the core of the cold-atom source: a 2D MOT (right) loads a 3D MOT (center) from which the atoms are launched vertically. On their way up, they pass through various light beams to prepare their quantum state (see text). The interferometer region doesn't appear on this picture. \label{fig:overview}} 
\end{figure}

The  Matter-wave laser Interferometric Gravitation Antenna (MIGA) aims at providing the first platform to study the key techniques for GW detection with matter wave interferometry. The project is described in details in Ref.~\cite{Canuel2018}. An overview of the instrument is also shown on figure \ref{fig:overview}. It will consist of two perpendicular arms of 150-meter long vacuum tubes (diameter of 50~cm) containing two parallel laser beams propagating at different heights (separated by 30~cm) in order to drive a cold-atom interferometer that measures the local, horizontal, component of the gravitational field. Three Bragg atom interferometers (AIs) will be distributed in each arm with a mutual interval distance of nearly 75~m. The differential measurements of gravity on various baselines will allow to extract the local gravity gradient.

The characteristics of the MIGA instrument (horizontal interrogation direction, Bragg diffraction) require a specific development for the atomic preparation and detection. An important feature of the instrument is the ability to produce cold atom clouds with the appropriate quantum state which are injected in the AI region, and to detect the state of the atoms at the output of the interferometer. A Bragg interferometer requires a narrow velocity selection for the input state, and the ability to detect the output velocity state. This article presents the working principle of the cold-atom systems (excluding the interferometer part), their design and realization, and their performances.

\section{Overview of the future MIGA instrument}

The future system relies on horizontal Bragg interferometers where three light pulses are used to split, deflect and recombine the atomic waves. The cold atoms enter the interferometer from the bottom and interact with two horizontal pairs of counterpropagating Bragg lasers separated by a distance of 30.6~cm in the vertical axis. The splitting and recombination ($\pi/2$) pulses are done with the bottom beam while the deflection ($\pi$) pulse is performed with the top beam. The separation between the 2 beams corresponds to an interferometer with a pulse separation time $T=250$~ms. The Bragg transitions at each position are based on the interaction of the atom with two counter-propagating laser beams of the same frequency, which couple the momentum states $|\vec{p}-\hbar\vec{k}\rangle$ and $|\vec{p}+\hbar\vec{k}\rangle$, where $\vec{p}$ is the momentum of the atom and $\vec{k}=\frac{2\pi}{\lambda} \hat{u}$ is the wave vector of the lasers ($\lambda\simeq 780$~nm, $\hat{u}$ unit vector of the laser propagation). With a vertical launch velocity $v_0=4.45$ m$/$s, the time spent from the launch to the apogee ($\pi$ Bragg pulse) is $t_{ap}=454$~ms.

The experimental sequence starts with a cold atom production phase during which the Rubidium 87 atoms are collected in a magneto-optical trap (MOT), laser cooled using polarization gradient cooling and vertically launched. On their way up before the interferometer, the atoms are prepared in the desired quantum state (internal state and velocity). On the way down, the probability for an atom to be in a given output momentum state is detected using a momentum-to-internal state labelling technique based on a Raman transition, followed by a fluorescence detection.

\section{Description of the cold atom head}

\subsection{The cold atom production}

\begin{figure}[h!]
\centering
\includegraphics[width=0.8\linewidth]{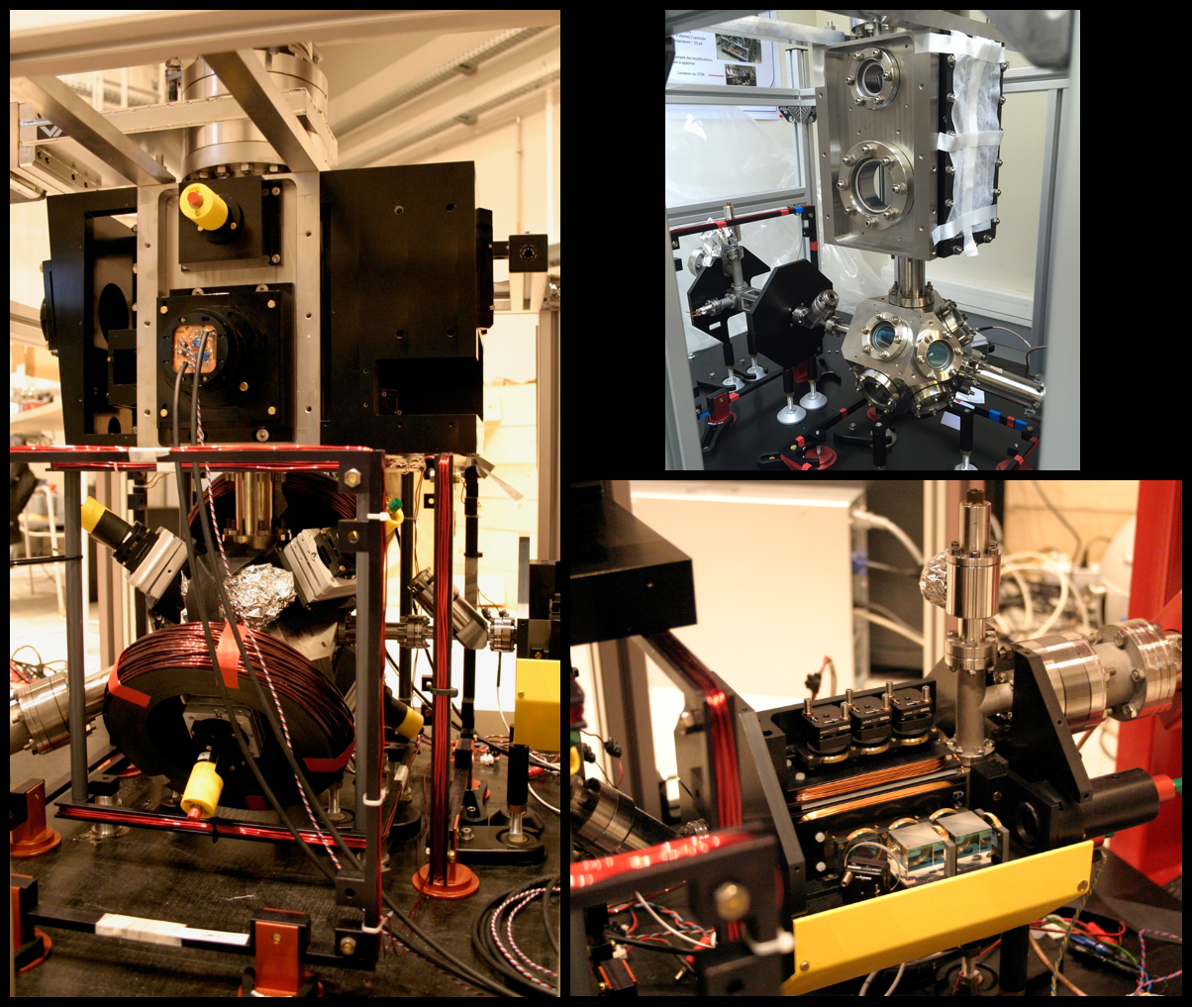}
\caption{ Photos of the cold atoms unit. Left: MOT and preparation/detection chamber with optics and magnetic coils. Top right: the complete vacuum system of the atom head without optics. Bottom right: The 2D MOT chamber with optics and magnetic coils.  \label{fig:3dmot_detection_chambers}} 
\end{figure}

A two-dimensional magneto-optical trap (2D MOT) fed by a rubidium alloy dispenser is realized in a vacuum chamber connected to the atom preparation chamber by a differential vacuum tube (see figure \ref{fig:overview} (b) and \ref{fig:3dmot_detection_chambers}), allowing to reach a loading rate of $6 \times 10^{8}$ atoms$/$s in a three-dimensional MOT (3D MOT).

The atoms from the 3D MOT are then simultaneously cooled down and vertically accelerated to a velocity of $v_0=4.45$ m$/$s in a moving molasses. Two acousto-optic modulators respectively driving three top and three bottom MOT beams allow to apply a detuning $\delta$ between each pair of counterpropagating beams. The three pairs of beams are mutually perpendicular and form a angle of $arccos(\frac{1}{\sqrt{3}})$ with the vertical axis, yielding an equilibrium velocity in the molasses of $v_0=\frac{\sqrt{3}}{2} \delta \lambda $, where $\lambda$ is the laser wavelength. After shutting off the MOT magnetic field, the global laser detuning from atomic transition is increased from 12 to 150 MHz and the laser intensity is decreased to zero in a 2 ms ramp, allowing the atomic cloud to reach a temperature of $\approx 2$ $\mu$K. The molasses duration is limited to a few ms by the transit time of the accelerated cloud through the MOT beams.

\subsection{The laser system}

The laser system used for the cold-atom sources is described in \cite{sabulsky2019fibered}. We give a short description here for consistency.
The different lasers used to cool and manipulate the atoms are delivered from an all-fibered laser module. The laser architecture is based on frequency doubled telecom lasers. A Master laser is locked using a  Rubidium 85 saturated absorption spectroscopy signal and references 3 slave lasers which are respectively used for the 2D MOT cooling laser, the 3D MOT cooling and Raman 2 laser, and the 3D MOT repumper and Raman 1 laser. The 3 slave lasers are all phase locked to the Master laser via an adjustable frequency offset. The repumping light for the 2D MOT is generated by a fiber electro-optic phase modulator at 1560~nm fed with the appropriate microwave frequency.

After amplification of the 1560~nm light emitted by laser diodes in Erbium doped fiber amplifiers and second harmonic generation in PPLN waveguide cristals, the 780~nm light is sent to optical splitters and guided to the experiment chamber using several optical fibers. The laser module nominally delivers 300~mW total power for the 2D MOT, 150~mW total power for the 3D MOT, 350~mW in the Raman 1 beam and up to 700~mW in the Raman 2 beam. The relative power and polarization fluctuations at the fiber outputs have been measured to be below  one percent rms on a timescale of 100 hours.
The phase lock signals are controlled by radio and microwave frequency sources all referenced to a stable 100 MHz quartz oscillator.
The full laser system is hosted in a $1.7\times 0.5\times 0.5 \ \text{m}^3$ transportable rack.
The control of the laser system and of the experiment is provided by home-made electronics \cite{BertoldiControlSystem}.

\subsection{The magnetic shield and bias field}

\begin{figure}[h!]
\centering
\includegraphics[width=0.6\linewidth]{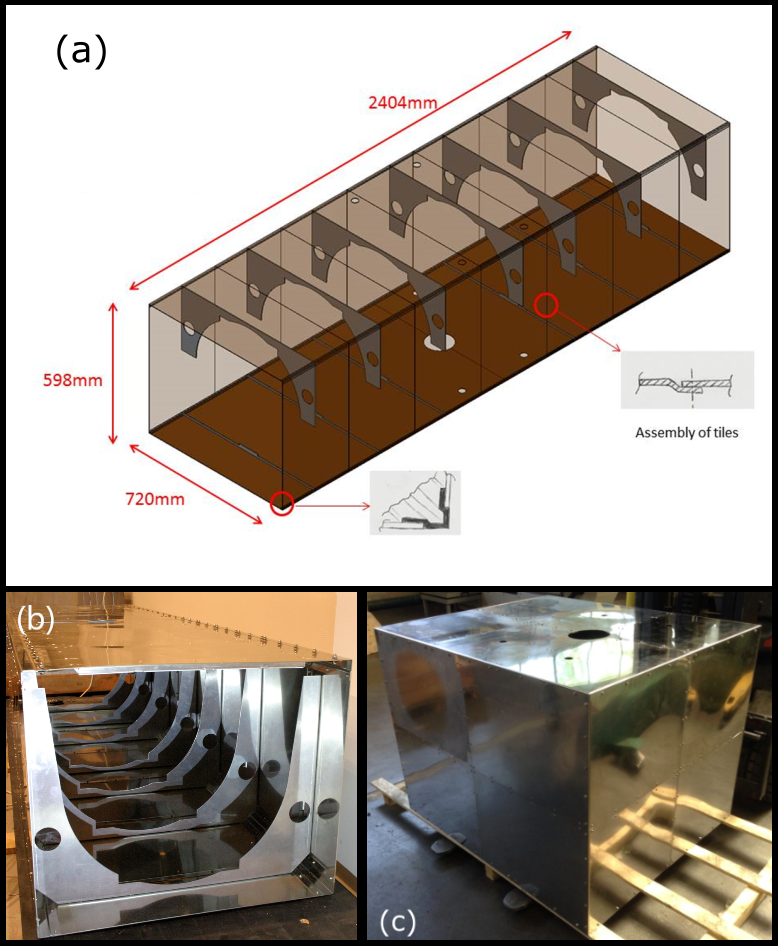}
\caption{Magnetic shields. (a)  Technical drawing of the outer shield surrounding a part of the vacuum tube above the cold-atom source. (b) Photograph of the shield with an open part. (c) Inner shield surrounding the cold-atom source \label{fig:magnetic_shield}} 
\end{figure}

The three AI parts of the instrument will be enclosed in mu-metal in order to maintain the stray magnetic field to a low level (see figure \ref{fig:magnetic_shield}). The interferometer regions will be enclosed in a $2404 \times 598 \times 720 $ mm magnetic shield in order to suppress the measurement bias due to the second order Zeeman effect. The atoms preparation region is also shielded in order to keep a low magnetic field during the moving molasses sequence. After turning off the MOT magnetic field, residual magnetic field at the position of the 3D MOT of $\approx 7$ mG was measured in-situ by microwave spectrocopy of the atomic hyperfine transition. 

A homogeneous horizontal bias magnetic field  of $\approx 200$ mG is produced by 4 vertical copper rods disposed around the atom head in order to define a quantization axis during the state preparation phase and the detection phase.

\subsection{Preparation of the atoms before the interferometer}

\begin{figure}[h!]
\centering
\includegraphics[width=0.8\linewidth]{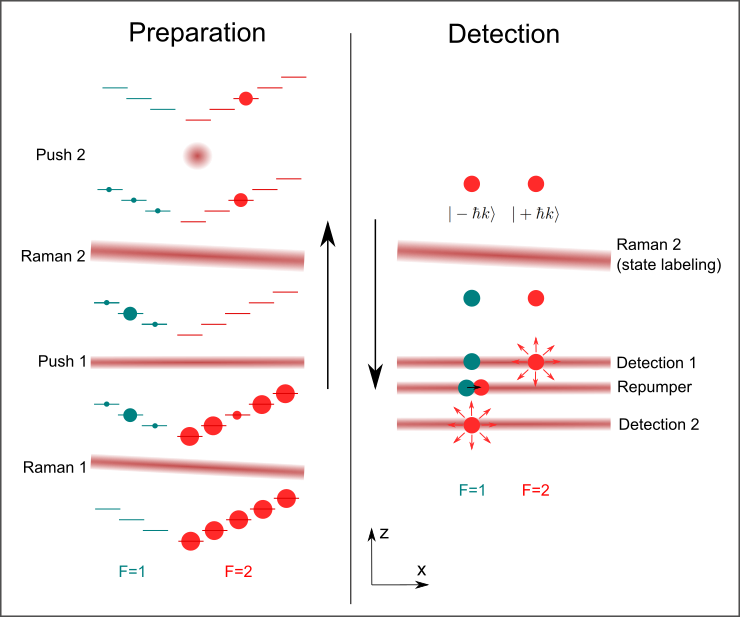}
\caption{ Atom preparation (left) and detection system (right). In the preparation, the atom cloud moving upward and initially containing atoms in all sublevels of $F=2$ goes through a series of Raman transitions and pushing beams, which leaves the remaining atoms in the $F=2,m_{F}=0$ state with a much reduced velocity spread in the x direction. For the detection, atoms are leaving the interferometer in a superposition of two momentum states. The population in one of them is transferred to $F=1$ with a Doppler sensitive Raman transition. State-selective fluorescence detection then allows to retrieve the output state and thus the interferometric phase. 
\label{fig:principle_preparation_detection}} 
\end{figure}

In order to inject the atoms with the right velocity in the Bragg interferometer, the quantum state (momentum and spin state) of the atoms is prepared on their way up, before the interrogation region. 
A first counter-propagating velocity-selective Raman pulse (beam 'Raman 1' in Fig.~\ref{fig:principle_preparation_detection}) is used to select the atoms from a mix of the Zeeman sub-levels of $|F=2\rangle$ to the $m_F=0$ Zeeman sub-level of the $|F=1\rangle$ hyperfine state, with a relatively narrow velocity class in the direction of the Raman lasers (e.g. corresponding to a temperature of $\approx 100$ nK). 
The  atoms which are not transferred to $|F=1\rangle$ by the Raman transition are then pushed by a laser tuned on resonance with the cycling transition ('push 1' in Fig.~\ref{fig:principle_preparation_detection}).  
A similar Raman/push procedure is then applied a second time to clean the remaining unwanted $F=1$ atoms produced by spontaneous emission during the first Raman selection pulse. For this purpose, we use a second Raman light pulse ('Raman 2', $1/e^{2}$ beam waist of $w=30$ mm) to transfer the atoms back to the $F=2$ state. The remaining atoms in the $F=1$ state are pushed with an orthogonal beam tuned on the $F=1\rightarrow F^\prime= 0$ transition ('push 2').

The angle of the Raman beams with respect to the horizontal plane is set to few degrees in order to introduce a Doppler effect and thus lift the degeneracy between the $|p\rangle\rightarrow |p+2\hbar k\rangle$ and $|p\rangle\rightarrow |p-2\hbar k\rangle$ transitions. The atoms must enter the interrogation region with a well defined velocity in order to fulfill the resonance condition in the  Bragg interferometer where the laser beams are horizontal (wavevector  $\vec{k}=k\vec{e}_x$). For a $2n\hbar k$  transition coupling the momentum states $|\vec{p}_0\rangle \rightarrow |\vec{p}_0+2n\hbar \vec{k}\rangle$, the input momentum   must fulfill $p_{0x}=-n\hbar k$. The vertical velocity at the first Bragg pulse is $gT\simeq 2.5$~m.s$^{-1}$, resulting in an angle $\theta=n\hbar k/MgT\simeq n\times 3.2$~mrad of the atom trajectory with respect to gravity. A rough control of this angle can be obtained by adjusting the directions of the 3D MOT Raman beams. Then, a more controlled step consists in selecting the right horizontal velocity class with the two step Raman selection by using different two-photon detunings for  the 'Raman 1' and 'Raman 2' beams.

\subsection{Detection of the atomic state}

\begin{figure}[h!]
\centering
\includegraphics[width=0.6\linewidth]{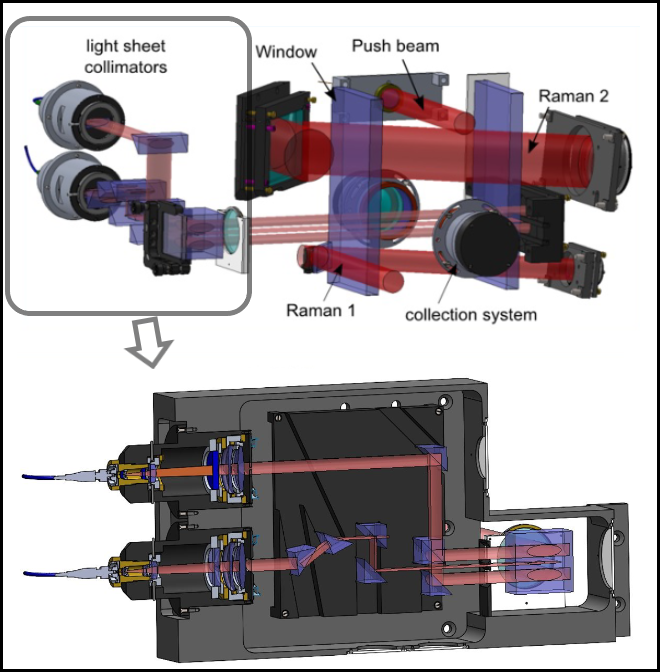}
\caption{ CAD drawing of the atomic preparation and detection system. The bottom panel shows how the 3 elliptical light sheets used for the detection are produced from the light originating from two elliptical collimators.   \label{fig:raman_detection_system}} 
\end{figure}

After their interrogation by the Bragg beams in the interferometer, the two different momentum states $|\pm n\hbar k\rangle$ of the atoms are
labeled to two different internal states with the Raman 2 laser beam (see figure \ref{fig:principle_preparation_detection}). More precisely, the velocity selective feature of the Raman transition is used to
transfer the $|F=2,- n\hbar k\rangle$ atoms to the $F=1$ internal state, while the $|F=2,n\hbar k\rangle$ atoms remain in the $F=2$ internal state. The
internal state of the atoms can then be resolved by state selective fluorescence detection. The falling atomic cloud crosses a series of horizontal light sheets.  Detection of the population in $F=2$ (that corresponds to the $|F=2,n\hbar k\rangle$ output state of the interferometer) is first realized by measuring the fluorescence induced by a first light sheet beam tuned on resonance with the $F = 2 \rightarrow F^\prime= 3$ transition. The retro-reflected light sheet is partially blocked at the
reflection mirror so that the $F=2$ atoms are pushed away in the process. 
The $F = 1$ atoms are then re-pumped to the $F = 2$ state using a thinner intermediate light sheet tuned to the $F = 1 \rightarrow F^\prime= 2$ transition. Those atoms are then detected in a third light sheet similar to the first one, measuring the population initially in $F=1$ (that corresponds to the $|F=2,- n\hbar k\rangle$ output state of the interferometer). The preparation and detection systems CAD drawings are shown in figure \ref{fig:raman_detection_system}.

\begin{figure}[h!]
\centering
\includegraphics[width=0.8\linewidth]{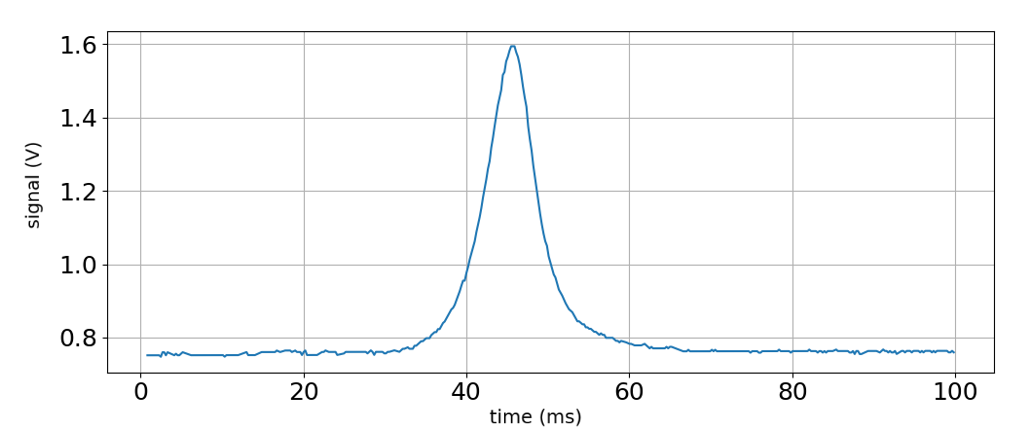}
\caption{ Time of flight fluorescence signal of the upper light sheet (atoms in $F=2$) recorded by the detection system. 
\label{fig:detectionstat}} 
\end{figure}

 The fluorescence light emitted by the atoms in the two light sheets is collected by a lens with a $2 \%$ collection efficiency and imaged on a two-quadrant photodiode. Figure \ref{fig:detectionstat} shows a typical time of flight fluorescence signal. The atomic population in a hyperfine state is retrieved by integrating the signal in the corresponding photodiode. The transition probability is deduced from the signals from the two light sheets (see section 2.6) after correction of systematic cross-talks between the two optical systems. The signal to noise ratio of the detection system is limited by electronic noise on the photodiode due to its dark current. The contribution from the electronic noise is $\sigma_{elec} = 25$ $\mu V/\sqrt{Hz}$. From a mean atomic signal of $S \simeq 100 $ mV we deduce a maximal signal to noise ratio of about 1000.

\section{Characterization of the system}

\subsection{State preparation}

\begin{figure}[h!]
\centering
\includegraphics[width=0.8\linewidth]{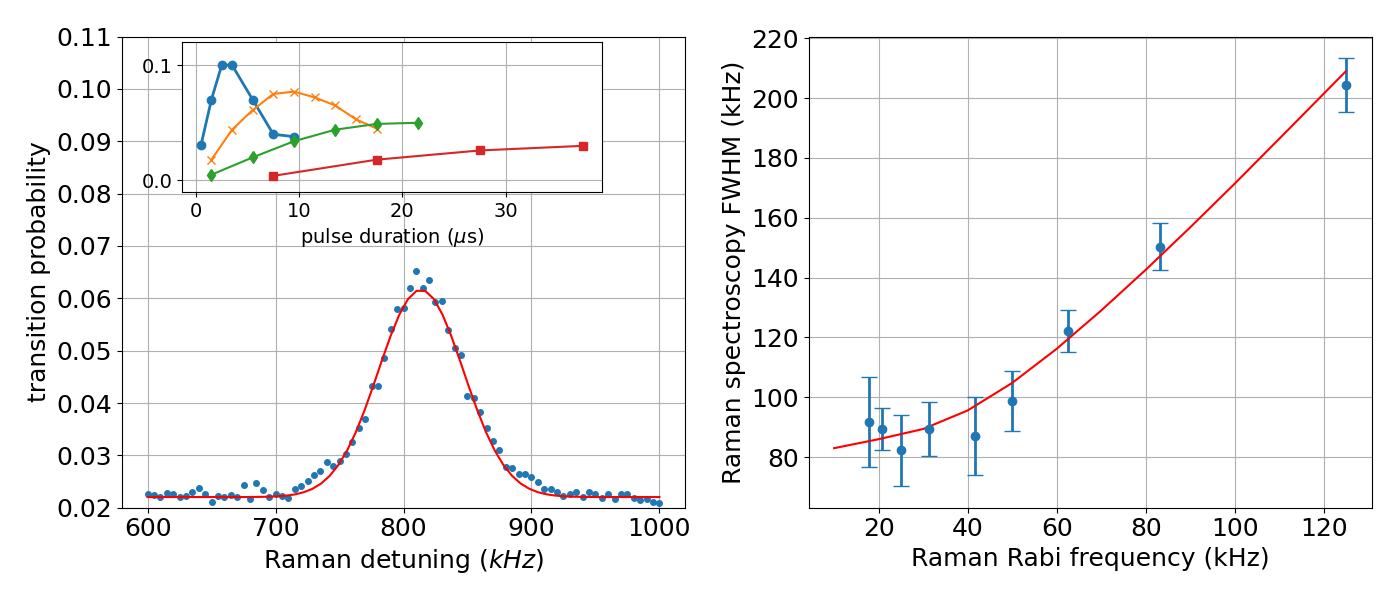}
\caption{Left: Raman spectroscopy. Transition probability as a function of Raman detuning, for a Rabi frequency of $\Omega= 2\pi \times 25$ kHz. The gaussian fit (solid line) indicates a full width at half maximum (FWHM) of $90$ kHz limited by the atomic velocity spread. Inset: Transition probability versus duration showing Rabi oscillations for different Raman light intensities.   Right: Raman spectrocopy FWHM of the peak for different Raman Rabi frequencies. Each point was extracted from a resonance peak obtained with a Raman pulse adjusted to half the Rabi period. The width of the peak is Fourier limited for large Rabi frequencies but limited to $ \approx 90$ kHz by velocity spread for lower values. The solid line shows the expected FWHM as the result of numerical integration of the convolution between a gaussian velocity distribution for a temperature $T=2$ $\mu$K and the profile of a square transition pulse.\label{fig:raman_spectro}} 
\end{figure}

In order to characterize the velocity selection during the state preparation phase, we performed Raman spectroscopy with the bottom Raman beam (Raman 1, see figure \ref{fig:raman_spectro}). Due to the Doppler effect, the observed resonance is at a Raman detuning of $ 810 $ kHz from the hyperfine transition. This corresponds to a component of the velocity of $0.31 $ m/s for the atoms along the Raman wave vector axis. Such value is consistent with an angle of 4.6 degrees between the Raman wave vector and the horizontal plane, given a vertical velocity for the atoms of $4.0$ m/s when they cross the Raman beam. The inset of figure \ref{fig:raman_spectro} (a) shows Rabi oscillations on the Raman 1 transition for different Raman light intensities. We observe a maximum transition probability of 10$\%$ for a Rabi frequency of $\Omega_{R}\approx 2\pi \times 100$ kHz. This is mostly limited by atoms from the molasses are distributed among all the different spin projection states and only a fraction of the atoms being initially in $F=2$, $m_F=0$. Moreover, transverse inhomogeneities of the light intensity lead to the damping of the oscillations and limit the contrast.

\begin{figure}[h!]
\centering
\includegraphics[width=0.7\linewidth]{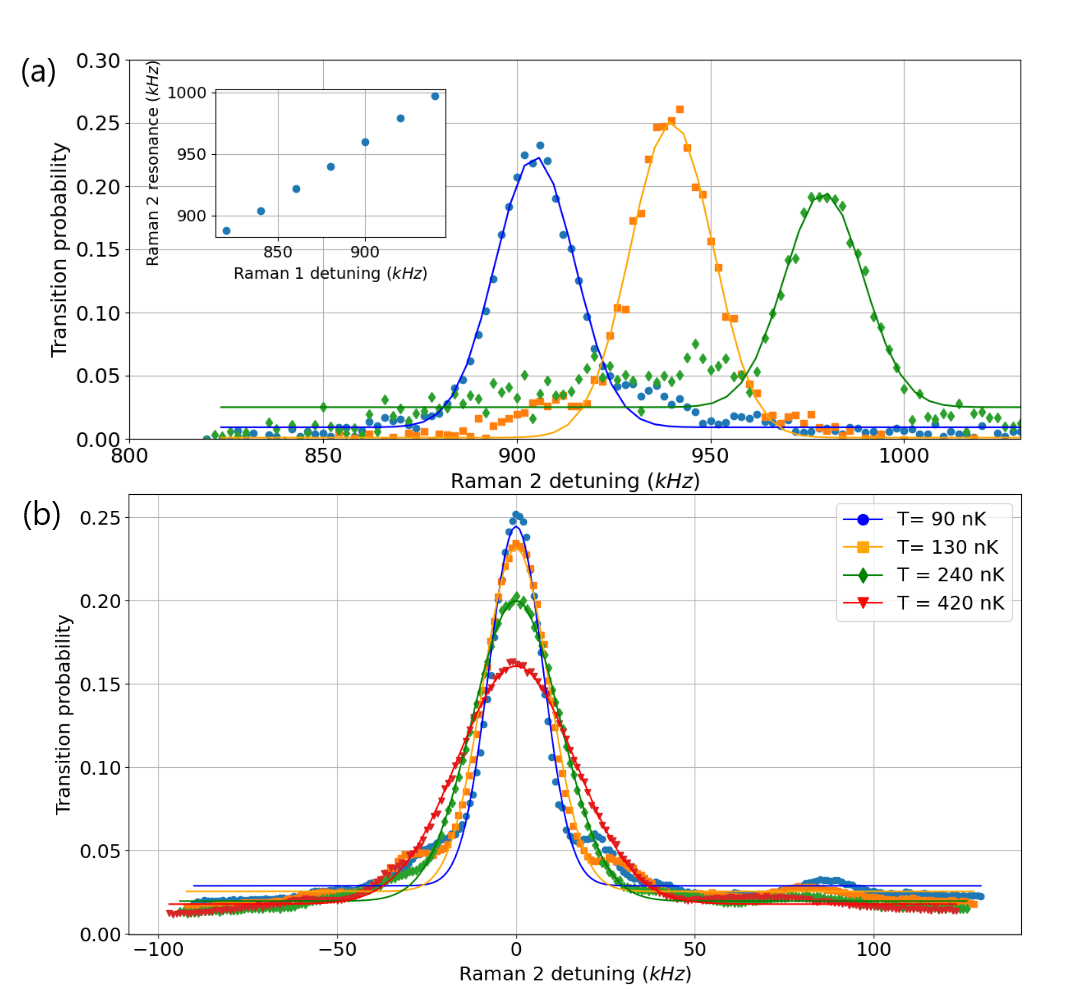}
\caption{(a) Raman spectroscopy performed with the upper Raman beam (Raman 2), for three different mean velocities selected with the lower Raman beam (Raman 1). The peaks correspond to three different Raman 1 frequencies incremented by 40 kHz steps. Inset: Raman 2 resonance as a function of Raman 1 frequency. The offset between the two frequencies is due to a difference in vertical velocity when the atoms cross the beam leading to a different Doppler shift. (b) Raman spectroscopy performed with the upper Raman beam (Raman 2), for four different velocity spreads selected with the lower Raman beam (Raman 1). The legend gives the corresponding effective temperatures deduced from a gaussian fit of the peak.
\label{fig:velocity_selection}} 
\end{figure}

Raman velocimetry allows to estimate the temperature of the cloud. We performed Raman spectroscopy for different Rabi frequencies, adjusting the pulse duration to half a Rabi period. Figure \ref{fig:raman_spectro} (b) shows the FWHM extracted from a Gaussian distribution adjustment as a function of the Rabi frequency. For large Rabi frequency the width of the peak is Fourier limited, whereas it reaches a constant value of $\approx 90$ kHz when the Rabi frequency is below $50$ kHz. This value is limited by the Doppler effect induced by the atomic velocity spread, which corresponds to an atomic temperature of $\approx 2$ $\mu$K.

The main function of the Raman 1 beam is to allow for a narrow horizontal velocity selection in order to couple the atoms to the Bragg interferometer with the right velocity. Figure \ref{fig:velocity_selection} (a) shows Raman spectroscopy performed with the top Raman beam (Raman 2), for different horizontal velocity distributions selected using Raman 1. For each peak we performed a Raman 1 transition with a different frequency, and a light intensity and pulse duration corresponding to a Fourier limited Gaussian FWHM of $\approx 23$ kHz in order to select a velocity spread corresponding to an effective temperature of $150$ nK along this direction. The spectroscopy was performed with a longer Raman 2 pulse in order to resolve the Doppler induced frequency width related to this temperature. Gaussian adjustments of the spectra are in good agreement with an effective temperature of $150$ nK. The  Figure \ref{fig:velocity_selection} (b) shows the same experiment, but for a single mean velocity selected and various effective temperatures. These results demonstrate that we have a good control on the momentum distribution that will enter the interferometer.

Figure \ref{fig:rabiRaman2} shows Rabi oscillations for the upper Raman transition, after velocity selection with the bottom Raman beam. That transition is the last step of the state preparation sequence, that allows to enter the Bragg interferometer in the state $F=2$, $m_F=0$ with a narrow velocity spread along the interferometer axis, and the right velocity for the first Bragg transition. In the process of selecting internal state and velocity, 98$\%$ of the atoms are removed and we obtain an effective flux of $1.2 \times 10^{7}$ atoms$/$s in the interferometer.

\subsection{Output state measurement}

\begin{figure}[h!]
\centering
\includegraphics[width=0.7\linewidth]{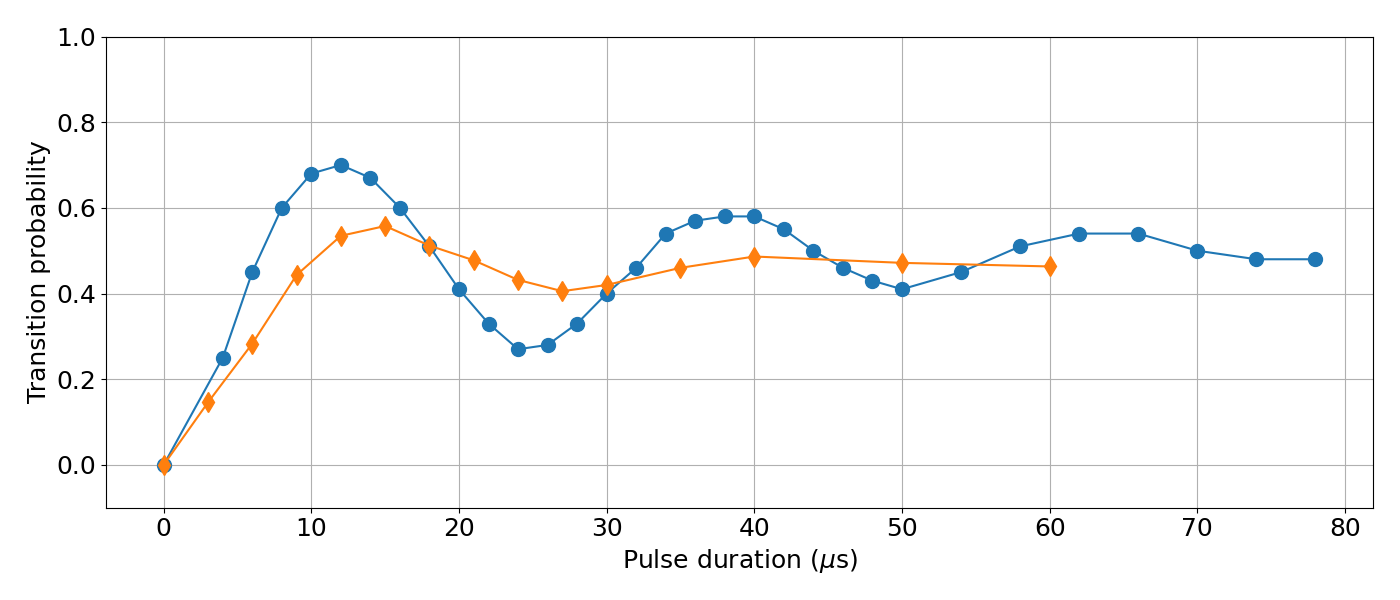}
\caption{ Transition probability versus pulse duration for Raman 2 transitions, performed when the atoms are going up (blue dots) and on their way down (orange diamonds). 
\label{fig:rabiRaman2}} 
\end{figure}

The measurement of the output state of the interferometer relies on the combination of external state labelling by a Raman transitions on the way down, and state selective fluorescence detection (see figure \ref{fig:principle_preparation_detection}). We evaluated the contrast of the Raman 2 transition on the way up and on the way down by recording Rabi oscillations visible on figure \ref{fig:rabiRaman2}. The selection pulse on the way up occurs at $T=80$ ms after launch, on the way down it occurs at $t=853$ ms. At the latter time we estimate the atom cloud diameter (FWHM of the atomic distribution) to be $3.3$ cm. We obtain a transfer efficiency of 55$\%$, limited by light intensity inhomogeneity across the laser beam due to the limited size of the gaussian profile of the beam (3.5 cm FWHM). The use of a top-hat laser profile is expected to increase this contrast to more than 70$\%$ \cite{Mielec2018}.  

\begin{figure}[h!]
\centering
\includegraphics[width=0.6\linewidth]{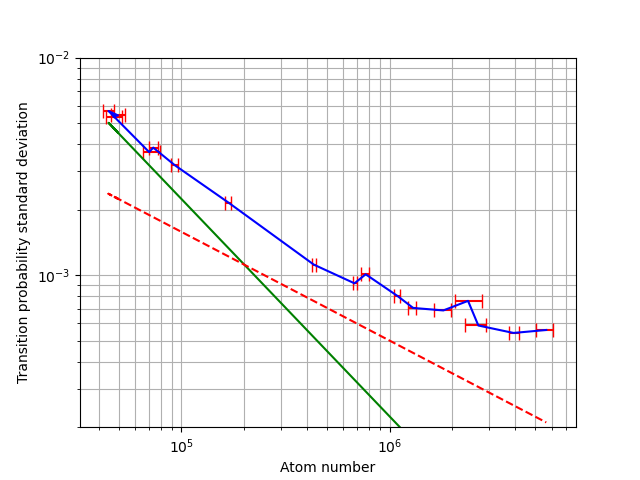}
\caption{Standard deviation of the measured transition probability, as a function of the number of atoms detected. The green line corresponds to $1/N$ the noise limit associated to the electronic noise from the photodiodes shown in figure \ref{fig:detectionstat}. The red dashed line corresponds to the quantum projection noise proportionnal to $1/\sqrt{N}$ . 
\label{fig:probatransstat}} 
\end{figure}

Finally, we investigated the detection noise by measuring the standard deviation of the measured transition probability, as a function of the number of atoms ($N$) detected, for a sample in a state of equal repartition between the two hyperfine levels (see figure \ref{fig:probatransstat}). This corresponds to the $1/2$ transition probability state often used as the output state of an interferometer. For low atom numbers, the noise follows a $1/N$ behaviour due to the electronic noise of the detection photodiodes. For high atom numbers, the noise saturates at a floor independent of the atom number, limited by fluctuations in the balance of the detection efficiency between the two hyperfine states. This noise floor level can be decreased by improving the power stability of the detection lasers.

\section{Conclusion}

We presented a characterization of the cold-atom sources developed for the MIGA interferometer. They allow the production of $10^{7}$ atoms per second selected in the desired velocity class for the Bragg interferometer, with a tunable effective temperature in the $100$ nK range along the interferometer axis. The interferometer's output state can be measured by a detection system comprising velocity selective Raman spectroscopy coupled to hyperfine state sensitive fluorescence detection. We reach a signal to noise ratio of 1000 with atoms in the desired state for interferometry, and a shot-to-shot standard deviation on the transition probability of $5\times 10^{-4}$, which fulfills the requirements for the future gravitation antenna. 

To date, five atom source units have been produced, three of which will be part of the 150 m long gradiometer in Rustrel and two others will be part of a $1.5$ m long gradiometer in LP2N in Bordeaux. More atom sources can be produced for potential extensions of the MIGA instrument (second horizontal axis, or extension of the baseline with more atom interferometers). Further characterisation and optimisation will be carried out during integration of the subsystem to the instrument. A possible design evolution may include the use of top-hat beam shapers to increase the contrast of Raman transitions \cite{Mielec2018}, adaptation of the detection system to large momentum transfer in the interferometer, or the use of interleaved interferometry \cite{Savoie2018}. 

\bibliography{Biblio}

\section*{Acknowledgements}

This work was realized with the financial support of the French State through the “Agence Nationale de la
Recherche” (ANR) within the framework of the “Investissement d’Avenir” programs Equipex MIGA (ANR-11-
EQPX-0028) and IdEx Bordeaux - LAPHIA (ANR-10-IDEX-03-02). This work was also supported by the region
d’Aquitaine (project IASIG-3D and USOFF). We also acknowledge support from the CPER LSBB2020 project;
funded by the “region PACA”, the “departement du Vaucluse”, and the “FEDER PA0000321 programmation
2014-2020”. We acknowledge the financial support from Ville de Paris (project HSENS-MWGRAV), DIM SIRTEQ, and Agence
Nationale pour la Recherche (project PIMAI, ANR-18-CE47-0002-01 and project EOSBCMR, ANR-18-CE91-0003-01). X.Z. thanks the China Scholarships Council (No 201806010364) program for financial support. J.J. thanks
“Association Nationale de la Recherche et de la Technologie” for financial support (No 2018/1565).

We acknowledge the contribution of Louis Amand to the design and assembly and of Bess Fang to preliminary tests of the system. We thank Jos\'{e} Pinto for his support with the electronic systems. 

\section*{Author contributions statement}

The project was designed by R.G. B.C. A.L. A.B and P.B. .The design of the atom heads was done by R.G, D.H, B.V., B.C. and A.L.. The systems were built and tested by R.G, Q.B, B.V., D.H., L.V.,M.L.,M.P., L.S., J.J., A.B., D.S., X.Z. and P.L. .The data were acquired and analysed by Q.B. and P.L. .The manuscript was written by Q.B and R.G. . All the authors have read, contributed to, and approved the final manuscript.

\end{document}